\documentclass[sn-mathphys-num]{sn-jnl}

\usepackage{graphicx}%
\usepackage{multirow}%
\usepackage{amsmath,amssymb,amsfonts}%
\usepackage{amsthm}%
\usepackage{mathrsfs}%
\usepackage[title]{appendix}%
\usepackage{xcolor}%
\usepackage{textcomp}%
\usepackage{manyfoot}%
\usepackage{booktabs}%
\usepackage{algorithm}%
\usepackage{algorithmicx}%
\usepackage{algpseudocode}%
\usepackage{listings}%
\usepackage{needspace}
\usepackage{subcaption}

\theoremstyle{thmstyleone}%
\theoremstyle{thmstyletwo}%

\theoremstyle{thmstylethree}%
\begin{document}

\title{Two-Laser Heterodyne Detection System for ALPS II}

\author*[1]{\fnm{Henry} \sur{Frädrich}}\email{henry.fraedrich@desy.de}

\author[1]{\fnm{Aaron D.} \sur{Spector}}

\author[1]{\fnm{Axel} \sur{Lindner}}

\author[2,3]{\fnm{Guido} \sur{Müller}}

\affil*[1]{ \orgname{Deutsches Elektronen-Synchrotron DESY}, \orgaddress{\street{Notkestr. 85}, \postcode{22607}, \state{Hamburg}, \country{Germany}}}

\affil[2]{ \orgname{Max-Planck-Institut für Gravitationsphysik (Albert-Einstein-Institut) and Leibniz Universität Hannover}, \orgaddress{ \postcode{30167}, \state{Hannover}, \country{Germany}}}

\affil[3]{\orgdiv{Department of Physics}, \orgname{ University of Florida}, \orgaddress{ \postcode{32611}, \city{Gainesville}, \state{Florida}, \country{USA}}}


\abstract{The Any Light Particle Search II employs a highly sensitive heterodyne detection system to measure an ultra-weak electromagnetic field that is reconverted from axions inside a regeneration cavity (RC). 
The initial design of the heterodyne system utilized a local oscillator laser (LO) that was resonant with the RC to form an interference beatnote with the weak signal field exiting the cavity. 
A weakness of this design is that it is possible for the out-of-mode field component of the LO to be larger than the in-mode component, allowing its noise to dominate the optical component of the noise budget and reducing the SNR of the experiment. 
This problem can be addressed by introducing a second, non-resonant laser as the LO. Since the in-mode light of this laser does not couple to the cavity,  significantly more of its power is reflected than when the LO is resonant. 
We present a model for the measurement noise that predicts these design changes increase the SNR by a factor of roughly 35. This model was then confirmed experimentally by injecting a weak signal to the RC and examining the difference in the sensitivity of the detection system in each of these configurations.
}

\keywords{Heterodyne Detection, Laser Interferometry}

\maketitle

\section{Introduction}

The Any Light Particle Search II (ALPS II) is an ongoing `light-shining-through-a-wall' experiment located at DESY in Hamburg, Germany \cite{alpsdesign}. ALPS II is searching for pseudo-Nambu-Goldstone bosons that lie beyond the Standard Model of particle physics, the most famous of which is the axion \cite{PhysRevLett.38.1440,PhysRevLett.40.223,PhysRevLett.40.279}. ALPS II uses postulated axion-photon mixing in a background magnetic field to convert photons to axions and subsequently reconvert them into measurable photons behind an opaque barrier that blocks the unconverted light. 
The re-conversion process is amplified by a high finesse optical cavity that resonantly enhances the signal power, called the Regeneration Cavity (RC) \cite{mueller,hoogeveen}. 
Given the expected low rates of both axion production and photon regeneration, ALPS II employs a highly sensitive heterodyne detection system that benefits from long integration times \cite{hallal}. 
This scheme uses a local oscillator laser reflected from the RC to form an interference beatnote ($P_{\mathrm{AC}}$) with the weak signal field leaving the cavity:
\begin{equation}
    P_{\mathrm{AC}}  =2S\cos(\Omega t+\theta) = 2\sqrt{\eta R_{\rm cav} n_{\mathrm{LO}} n_{S}}\cos(\Omega t+\theta).
\end{equation}
The local oscillator's power in photons per second is $n_{\mathrm{LO}}$, while the power of the weak signal field is $n_{S}$. 
The power reflectivity of the cavity for the fundamental mode on resonance is $R_{\rm cav}$, the power spatial overlap between the fields is $\eta$, while $\theta$ is the differential phase relationship between them that is maintained by the ALPS II optical system. 
Keeping $\theta$ fixed ensures that the beatnote is coherent with the detection system. 
The signal is demodulated at the known difference frequency $\Omega$ of the two fields, in both quadratures, because $\theta$ is unknown \cite{hallal}. The result gives a measurement of the amplitude of the beatnote $S$. This amplitude is then squared and calibrated to calculate the regenerated photon rate \cite{opticsresults}:
\begin{equation}
    n_S = \frac{S^2}{\eta R_{\rm cav} n_{\mathrm{LO}}}.
\end{equation}
The measurement of $S$ is limited by noise, whose variance takes the form
\begin{equation}
\sigma_S^2=\sigma_{\mathrm{LO}}^2+\sigma_{\mathrm{NEP}}^2.
\end{equation}
Here, $\sigma_{\mathrm{LO}}^2$ denotes the variance due to the photon counting statistics of the local oscillator measured in reflection of the cavity. For a measurement bandwidth $\Delta f = 1/\tau$, where $\tau$ is the integration time, the variance in the measurement of the amplitude is equal to $n_{\mathrm{LO}}/\tau$. Decomposing this power into the fundamental mode and higher-order mode contributions yields
\begin{equation}
    \sigma_{\mathrm{LO}}^2
    = n_{\mathrm{LO}}\Delta f
    = \sigma_{\mathrm{LO},00}^2+\sigma_{\mathrm{LO},\mathrm{HOM}}^2
    = (n_{\mathrm{LO},00}+n_{\mathrm{LO},\mathrm{HOM}})/\tau.
\end{equation}
The power of the field component that interferes with the signal mode and contributes to both the signal and the noise is given by  $n_{\mathrm{LO},00} = \eta R_{\rm cav} n_{\mathrm{LO}}$ (\textit{in-mode}). 
The power associated with the higher-order mode (HOM) content of the light  is denoted by $n_{\mathrm{LO},\mathrm{HOM}}=(1-\eta)n_{\mathrm{LO}}$ (\textit{out-of-mode}). It is comprised of all of the detected photons that do not contribute to the signal amplitude.
The term $\sigma_{\mathrm{NEP}}^2 = n_{\mathrm{NEP}} /\tau$ includes other sources of technical noise and is dominated by the electronic noise of the photodetector chain. 
Here $n_{\mathrm{NEP}}$ is the noise equivalent power of the technical noise in terms of photons per second. 

In ALPS II, the critical measured variable is the signal power $n_S$. Therefore, we define the signal-to-noise ratio (SNR) in terms of the beatnote amplitude squared, $S^2$, and the variance in the measurement of the beatnote amplitude, $\sigma_{S}^2$:
\begin{equation}\label{SNR}
    \mathrm{SNR} = \frac{S^2}{\sigma_S^2}.
\end{equation}
A quantum limited detection process, which maximizes the SNR, requires that $n_{\mathrm{LO},00}\gg n_{\mathrm{LO},\mathrm{HOM}},n_{\mathrm{NEP}}$ such that the noise of the in-mode part of the local oscillator dominates over the noise from all other contributing factors. 
The SNR in this case is $\mathrm{SNR} = n_S \tau$, which is equal to the total number of signal photons measured over the integration time.

In the case of ALPS II, a laser is injected into the RC from the detector side and stabilized to a resonance using a Pound-Drever-Hall (PDH) lock \cite{hallal}. 
The reflected field serves as the local oscillator for the heterodyne detection system, while the transmitted field carries the cavity resonance information to the rest of the experiment. 
For given cavity losses, the signal field is maximized when the RC is impedance matched and the transmissivity of the output mirror matches all other cavity losses. 
However, in this case, the reflectivity of the cavity will be zero ($R_{\rm cav}=0$) and there will be no LO power to form an interference beatnote with the signal field ($n_{\mathrm{LO},00} = 0$).
When the cavity is nearly impedance matched, the in-mode terms in the noise and the signal-to-noise ratio go to zero ($n_{\mathrm{LO},00}, \rm{SNR} \rightarrow 0$).
This work motivates the introduction of an additional laser field that is fully reflected from the RC and functions as a dedicated local oscillator, independent of the impedance matching of the cavity. 

\section{Signal-to-Noise Model}
\label{Sec:SNR_model}
The $123\,\mathrm{m}$ long RC and its mirrors M1 and M2 are enclosed in a vacuum system that spans from the Central Station clean room, where M2 is mounted, to the Detector End Station, where M1 is mounted. 
As the axion field passes through the magnetic field inside the RC, it has an electromagnetic field component that is stored inside the cavity. The power of this field is resonantly enhanced by the cavity power gain and exits the cavity on both sides, attenuated by the cavity mirror transmissivities. The resonant enhancement of the signal power exiting the cavity at the detector side can be approximated by
\begin{equation} \label{beta}
    \beta \approx \frac{4T_{1}}{A^2},
\end{equation}
where $T_{1}$ is the transmissivity of M1, and $A = T_{1}+T_{2} +L$ is the total round-trip attenuation of the RC. $T_{2}$ is the transmissivity of M2 and $L$ represents all other optical losses inside the cavity. 
For given losses $L$, the highest achievable $\beta$ occurs when the cavity is impedance matched and $T_1 = L+ T_2$ \cite{todd}.

The RC's power reflectivity can be approximated by
\begin{equation}\label{eq::Rcav}
    R_{\mathrm{cav}} \approx \left(1-\frac{2T_1}{A}\right)^2. 
\end{equation}
This equation specifically denotes the reflectivity of the in-mode component of a laser field frequency stabilized to a cavity resonance. The result of the destructive interference between the promptly reflected field and the cavity leakage field at the input coupling  mirror can lead to the cavity having a lower reflectivity when nearly impedance matched \cite{opticsresults,todd}. 
This can be seen in the equation as $R_{\mathrm{cav}}\rightarrow0$ when $T_1 \rightarrow L+ T_2$.

In Section~\ref{Sec:1L} we describe a model of the SNR based on Equation~\ref{SNR} for a setup using a single injected laser as the LO. Then, in Section~\ref{Sec:2L}, the SNR of the system using a second, non-resonant LO is modeled. In Section~\ref{Sec:comp} the results of the models are compared.

\subsection{One Laser Setup}
\label{Sec:1L}
For a heterodyne detection system that uses a single laser injected into the cavity, the signal field leaking out of the cavity through M1 interferes with the in-mode part of the injected laser that is reflected at M1. This results in a beat-note with an amplitude $S_{1L}$, at the heterodyne frequency $\Omega_{1L}$, a fixed random phase $\theta$, and a noise variance of $\sigma_{1L}^2$. The SNR for this measurement of $n_S$ then takes the form:
\begin{align}\label{1L}
    \mathrm{SNR}_{1L} &= \frac{S_{\mathrm{1L}}^2}{\sigma_{\mathrm{1L}}^2}\\
    &= \frac{S_{\mathrm{1L}}^2}{\sigma_{\mathrm{inj},00}^2 + \sigma_{\mathrm{inj,HOM}}^2+\sigma_{\mathrm{NEP}}^2}\\
    &= \frac{n_{ \mathrm{inj}}\eta R_{\mathrm{cav}} \beta \tau}{n_{\mathrm{inj}}\eta R_{\mathrm{cav}} +n_{\mathrm{inj}}(1-\eta)+n_{\mathrm{NEP}}}.
\end{align}
Here $\sigma_{\mathrm{inj}}^2=\sigma_{\mathrm{inj},00}^2+\sigma_{\mathrm{inj},\rm{HOM}}^2$ is the variance due to shot noise of the injection laser measured in reflection of the cavity, and $\eta$ is the power spatial overlap between the injected laser and the cavity eigenmode. 
When the cavity is impedance matched and $R_{\rm cav}$ is zero, the signal will vanish while the noise terms $\sigma_{\mathrm{inj,HOM}}^2$ and $\sigma_{\mathrm{NEP}}^2$ remain. Subsequently, the SNR goes to zero for an impedance matched cavity.

\subsection{Two Laser Setup}
\label{Sec:2L}
The degradation of the SNR when the cavity is close to impedance matching can be avoided by introducing a second laser to the detector side of the system, referred to as the reflected laser ($L_{\mathrm{refl}}$). 
This setup will be referred to as the two laser setup. 
$L_{\mathrm{refl}}$ is frequency stabilized to the resonant injected laser ($L_{\mathrm{inj}}$) with a phase-locked loop before the lasers are incident on the cavity. The difference frequency is chosen such that $L_{\mathrm{refl}}$ will be held off resonance  and will therefore be fully reflected and independent of the cavity. 

In the two laser setup, the in-mode component of $L_{\rm refl}$ forms a beatnote with the regenerated axion field leaking out of the cavity at the new heterodyne frequency $\Omega_{\mathrm{2L}}$. We define $\sigma_{\mathrm{refl}}^2 = n_{\mathrm{refl}}/\tau$ as the variance in the beatnote amplitude measurement due to shot noise generated by $L_{\mathrm{refl}}$ incident on the cavity. The SNR can then be calculated using the following equation:
\Needspace{6\baselineskip}
\begin{align}\label{2L}
    \mathrm{SNR}_{\mathrm{2L}} 
    &= \frac{S_{\mathrm{2L}}^2}{\sigma_{\mathrm{2L}}^2}\\
    &= \frac{S_{\mathrm{2L}}^2}{\sigma_{\mathrm{refl}}^2+ \sigma_{1L}^2}\\
    &= \frac{\eta n_{\mathrm{refl}} \beta \tau}{n_{\mathrm{refl}}+n_{\mathrm{inj}}\eta R_{\mathrm{cav}} + n_{\mathrm{inj}}(1-\eta)+n_{\mathrm{NEP}}}.
\end{align}
For an impedance matched cavity configuration, the signal is not attenuated by a low $R_{\mathrm{cav}}$ in this setup. The noise, however, behaves similarly to that of the one laser setup, with an additional shot-noise contribution arising from $n_{\mathrm{refl}}$.

\subsection{Sensitivity Improvement}
\label{Sec:comp}
The SNRs from equations \ref{1L} and \ref{2L} depend on the total roundtrip losses $A$, the power spatial overlap of the input field with the cavity eigenmode $\eta$, the cavity input mirror transmissivity $T_1$, the injected lasers power $n_{\mathrm{inj}}$, the reflected lasers power $n_{\mathrm{refl}}$, the noise equivalent power of the technical noise $n_{\mathrm{NEP}}$, and the total measurement time $\tau$. 
To properly assess the SNR of the one laser and two laser setups, the system was characterized and these parameters were measured with the system in a configuration similar to the first science campaign \cite{opticsresults}. 

To align the system, the $A$ was minimized and $\eta$ was maximized, attaining the highest $\beta$ and reducing the out-of-mode component $n_{\mathrm{inj,HOM}}$. 
The former was achieved by translating the RC's eigenmode across the mirror surfaces, aligning the cavity eigenmode to the mirror spots where losses were lowest.
The input field was then angularly and laterally aligned to the new eigenmode position using input alignment mirrors.
The measurement techniques used to acquire $A$ and $\eta$ are outlined in \cite{isogai} and \cite{todd}, with the latter also presenting a measurement of $T_1 = 100\pm 1\,\mathrm{ppm}$.
The powers of $L_{\mathrm{inj}}$ and $L_{\mathrm{refl}}$ are measured on a photodetector in reflection of the cavity, producing the voltages $V_{\mathrm{inj}}$ and $V_{\mathrm{refl}}$, which are directly proportional to the powers $n_{\mathrm{inj}}$ and $n_{\mathrm{refl}}$, respectively. 
The powers were selected to roughly represent the powers used in the first science run.
The acquired parameters can be found in Table \ref{table1}, including the value of $R_{\mathrm{cav}}$, calculated from $A_{\mathrm{meas}}$ using equation \ref{eq::Rcav}.

\begin{table}[]
    \centering
    \begin{tabular}{ccccc}
        \toprule
        $A_{\mathrm{meas}}[\mathrm{ppm}]$ & $\eta[\,-\,]$ & $V_{\mathrm{inj}} [\mathrm{V}]$ & $V_{\mathrm{refl}}[\mathrm{V}]$ & $R_{\mathrm{cav}} [\,-\,]$  \\ 
        \midrule
        $239(1)$ & $0.949(4)$ & $2.676(4)$
        & $2.528(10)$ & $0.0267(30)$ \\
        \botrule
    \end{tabular}
    \caption{\label{table1}
        Parameters relevant to the SNR model extracted under conditions with the system aligned and used for the comparison of the one laser and two laser setups. $A_{\mathrm{meas}}$ is the total round-trip attenuation of the RC measured via storage time measurements. $\eta$ is the power spatial overlap of the input fields to the RC. $V_{\mathrm{inj}}$ and $V_{\mathrm{refl}}$ are the individual direct current voltages of the fully reflected injected and reflected laser respectively.$R_{\mathrm{cav}}$ is the cavity power reflectivity calculated from \ref{eq::Rcav} at the total round-trip losses $A_{\mathrm{meas}}$.
    }
\end{table}

The data shows that the out-of-mode reflected field component, attenuated by $(1-\eta)$, is significantly larger than the in-mode reflected field component, which is attenuated by $\eta R_{\mathrm{cav}}$. Under these conditions, the noise in the one laser setup must be dominated by either shot noise due to optical power that doesn't contribute to the signal or technical noise.

\begin{figure}[ht]
    \centering
    \includegraphics[width=\textwidth]{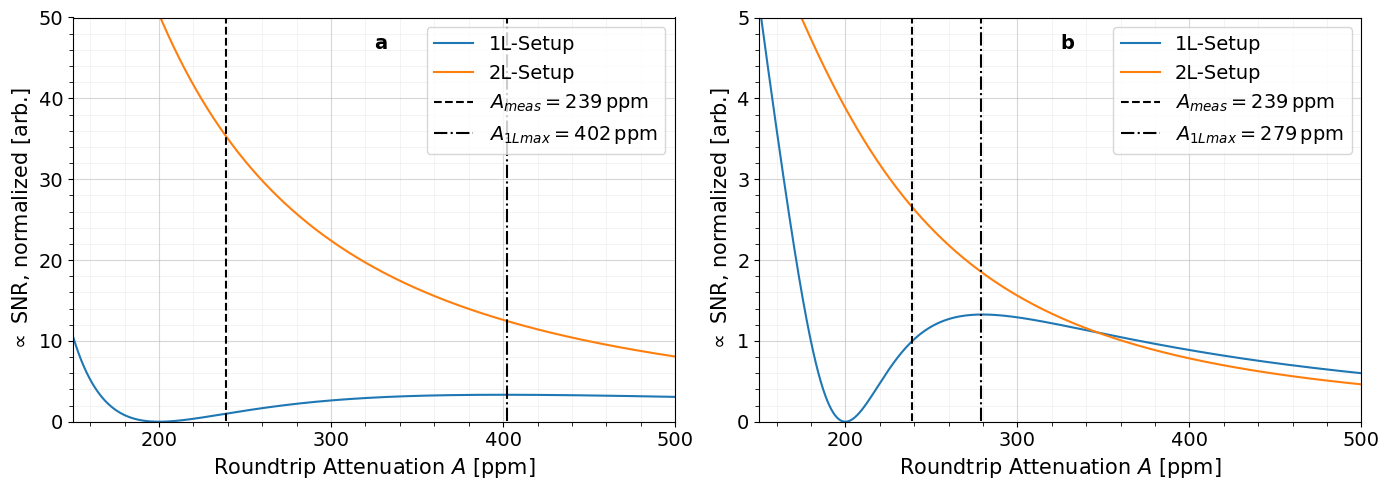}
    \caption{
    Comparison between modeled SNRs of the one and two laser setups (1L and 2L) as a function of the total round-trip attenuation of the RC in both detector chain electronic noise dominated case (a) and the shot noise dominated case (b). $A_{\mathrm{1Lmax}}$ and $A_{\mathrm{meas}}$ represent the total round-trip attenuation for the SNRs of the one and two laser setups respectively. }
    \label{fig:sim}
\end{figure}

The data was also used to calculate the normalized SNRs of both setups as a function of $A$. 
Here two cases were considered, both depicted in Fig. \ref{fig:sim}.
In the first case, the noise is dominated by electronic noise in the detector chain ($\sigma_{\mathrm{NEP}}^2 = \sigma_{\mathrm{1L}}^2,\sigma_{\mathrm{2L}}^2$). In the second case, the noise is dominated by optical shot noise  ($ \sigma_{\mathrm{1L}}^2, \sigma_{\mathrm{2L}}^2 \gg \sigma_{\mathrm{NEP}}^2$ ).
The first case represents the state of the system for the first science campaign, since the technical noise of the measurement was roughly a factor of six above the shot noise \cite{opticsresults}. 
The shot-noise-limited case is representative of the system after future upgrades to the detection electronics that will allow it to reach the shot-noise limit.
In both cases, both setups are normalized to the SNR of the one laser setup when the losses were tuned to the minimum value of $A_{\mathrm{meas}} = 239\,\mathrm{ppm}$. 
The blue and orange curves show the SNR predicted model as a function of the RC's total round-trip attenuation $A$ for the one and two laser setups, respectively.

Although $\beta$ increases with smaller $A$, in both cases, a reduction in $\mathrm{SNR}_{\mathrm{1L}}$ around the impedance matching condition can be observed and the SNR becomes zero for an impedance matched cavity. 
Within the operational range of the RC ($A\geq A_{\mathrm{meas}}$), $\mathrm{SNR}_{\mathrm{1L}}$ takes on its maximum values at $A_{\mathrm{1LmaxSN}} = 279\,\mathrm{ppm}$ for the shot noise dominated case and $A_{\mathrm{1LmaxTN}} = 402\,\mathrm{ppm}$ for the electronic noise dominated case. In both plots in Fig. \ref{fig:sim}, $A_{\mathrm{meas}}$, $A_{\mathrm{1LmaxSN}}$ and $A_{\mathrm{1LmaxTN}}$ are marked by vertical dashed and dash-dotted lines. 

We define the SNR ratio between the two setups as $G$ in the following equation:
\begin{equation}
    G(A_{\mathrm{1L}},A_{\mathrm{2L}}) = \frac{\mathrm{SNR}_{2L}(A_{\mathrm{2L}})}{\mathrm{SNR}_{1L}(A_{\mathrm{1L}})}.
\end{equation}
The case limited by technical noise yields:
\begin{equation}\label{Gdn}
    G_{\mathrm{TN}}(A_{\mathrm{1L}},A_{\mathrm{2L}})
    = \frac{V_{\mathrm{refl}} \beta(A_{\mathrm{2L}})}{R_{\mathrm{cav}}(A_{\mathrm{1L}}) V_{\mathrm{inj}} \beta(A_{\mathrm{1L}})}.
\end{equation}
If $A_{\mathrm{1L}} = A_{\mathrm{2L}}$ and $V_{\mathrm{inj}} = V_{\mathrm{refl}}$, the expression simplifies further, with $G_{\mathrm{TN}} = 1/R_{\mathrm{cav}}$.
The shot noise limited case yields
\begin{align}\label{Gsn}
    G_{\mathrm{SN}}(A_{\mathrm{1L}},A_{\mathrm{2L}})
    &= \frac{V_{\mathrm{refl}} \beta(A_{\mathrm{2L}})}{R_{\mathrm{cav}}(A_{\mathrm{1L}}) V_{\mathrm{inj}} \beta(A_{\mathrm{1L}})} \frac{V_{\mathrm{inj}}\eta R_{\mathrm{cav}}(A_{\mathrm{1L}})+V_{\mathrm{inj}}(1-\eta)}{V_{\mathrm{refl}}+V_{\mathrm{inj}}\eta R_{\mathrm{cav}}(A_{\mathrm{2L}})+V_{\mathrm{inj}}(1-\eta)}.
\end{align}
At $A_{\mathrm{meas}}$, technical noise dominated case yields $G_{\mathrm{TN}}(A_{\mathrm{meas}},A_{\mathrm{meas}}) = 35.4\pm4$, while the case dominated by only shot noise yields $G_{SN}(A_{\mathrm{meas}},A_{\mathrm{meas}}) = 2.65\pm0.22$. 
Although the improvement in the shot-noise-limited case is relatively modest, this result should be considered in the context of planned future upgrades. In addition to achieving shot-noise-limited detection, these upgrades will aim to maximize $\beta$ through improvements to the cavity optics and to shift the cavity operating point closer to impedance matching.

Considering the maximum values for $\mathrm{SNR}_{\mathrm{1L}}$ at $A_{\mathrm{1LmaxSN}}$ and $A_{\mathrm{1LmaxTN}}$ leads to $G_{\mathrm{TN}}(A_{\mathrm{meas}},A_{\mathrm{1LmaxTN}}) = 10.7\pm1.4$ and $G_{\mathrm{SN}}(A_{\mathrm{meas}},A_{\mathrm{1LmaxSN}}) = 2.00\pm0.04$.

\section{Experimental Methods}\label{meas}

\begin{figure}
\centering
\includegraphics[width=\textwidth]{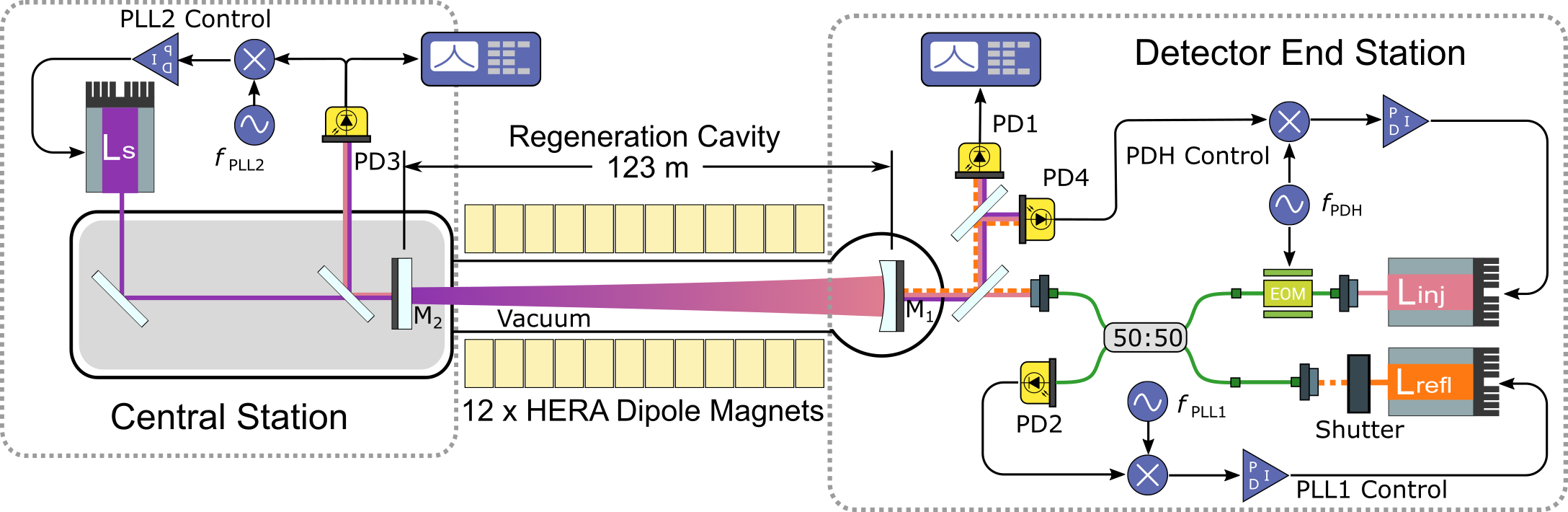}
\caption{Experimental setup used to measure the sensitivity improvement between the one and two laser configurations of the detection system.}
\label{fig::layout}
\end{figure}

A direct measurement of $G(A_{\mathrm{meas}},A_{\mathrm{meas}})$ was performed to assess the accuracy of the model.
The experimental layout is shown in fig.~\ref{fig::layout}. 
A small calibration signal emulating the electromagnetic field regenerated in the RC was injected from the Central Station into the RC to compare the SNRs of both setups.

The experiment utilizes three lasers in total: $L_{\mathrm{inj}}$ and $L_{\mathrm{refl}}$, which are located at the Detector End Station, and $L_{S}$, which provides the calibration signal from the Central Station.
$L_S$ is attenuated to approximately $10\,\mathrm{nW}$ before reaching the cavity, which further reduces the power to about $200\,\mathrm{pW}$ at the detector.
$L_{\mathrm{inj}}$ and $L_{\mathrm{refl}}$ are interfered in a fiber setup containing a 50:50 beam combiner with two outputs.
For the one laser setup, $L_{\mathrm{refl}}$ is blocked.
The light from one output port exits the fiber through a collimator and travels to the RC. 
The other output is connected to a photodetector (PD2), which is used to frequency-stabilize $L_{\mathrm{refl}}$ to $L_{\mathrm{inj}}$ at the off-resonance frequency $f_{\mathrm{PLL1}} = 54.95\,\mathrm{MHz}$, utilizing a phase-locked loop (PLL1). 
The fiber setup also contains an electro-optic modulator in the input path of $L_{\mathrm{inj}}$, which produces the necessary sidebands for the frequency control loop that stabilizes it to the RC \cite{pdh}.
The retro-reflected light from the RC is extracted from the beam path using a Faraday isolator.
A fraction of that light is picked off to PD4, which is used for the frequency stabilization setup.
The remaining light is incident on PD1. 

With $L_{\mathrm{inj}}$ stabilized on an RC resonance, its transmitted field exiting the cavity at the Central Station overlaps with $L_{S}$ at M2. 
The resulting beatnote is measured at PD3 and is used to stabilize $L_{S}$ to a cavity resonance with a frequency difference from $L_{\mathrm{inj}}$ of $f_{\mathrm{PLL2}} = m \cdot \mathrm{FSR}$. Here, the free spectral range of the RC $\mathrm{FSR} \approx 1.22\,\mathrm{MHz}$ and $m$ is an integer value.
For the one laser setup, $m=12$ corresponds to $f_{\mathrm{PLL2}} \approx 14.67\,\mathrm{MHz}$. 
This value was chosen because the beatnote of interest on PD1 will be at nearly the same frequency, $f_{\mathrm{HET}} = 14.67\,\mathrm{MHz}$, as the heterodyne frequency used during the first science campaign \cite{opticsresults}. 
To achieve a similar heterodyne frequency for the two laser setup, the locking frequency for PLL2 was changed to $f_{\mathrm{PLL2}} = 33\cdot \mathrm{FSR}$, leading to a heterodyne frequency at PD1 of $f_{\mathrm{HET}} = f_{\mathrm{PLL1}} - 33\cdot \mathrm{FSR} \approx14.60\,\mathrm{MHz}$ between $L_{S}$ and $L_{\mathrm{refl}}$.
For both setups, the beatnote amplitude at their respective $f_{\mathrm{HET}}$ is recorded on a spectrum analyzer.

In tandem with the signal measurements, noise levels were acquired by purposefully tuning $f_{\mathrm{PLL2}}$ to an off-resonance value while maintaining the same frequency range. 
To compare measurements, the same bandwidth of $\Delta f=48.87\,\mathrm{Hz}$ was maintained throughout, selected to be small enough to ensure adequate beatnote visibility while large enough to reduce measurement times.
Comparing both setups' measured beatnote amplitudes and noise levels at PD1 at the respective heterodyne frequencies provides a calibrated signal measurement that can be used to compare the setups. 

To ensure comparable conditions, both measurements were performed consecutively within a short time interval. 
Key parameters, including the values listed in Table \ref{table1}, as well as potential clipping losses and photodetector gains, were measured before and after the runs to verify the absence of significant system drifts and to assess the uncertainty of the measurement. 

The power of $L_{S}$ was monitored on PD3, exploiting its interference with the field of $L_{\mathrm{inj}}$ transmitted through the RC. 
Direct tracking of $L_S$'s power was not possible due to very low power levels.
Furthermore, a direct measurement of $R_{\mathrm{cav}}(A_{\mathrm{meas}})$ is performed by comparing the beatnote amplitudes between $L_S$ and $L_{\mathrm{refl}}$ for $L_{\mathrm{refl}}$ stabilized on and off resonance of the RC.

\section{Results}
\label{Sec:Results}
The measured SNR ratio between the one laser and two laser setups is:

\begin{equation}
    G_{\mathrm{meas}}(A_{\mathrm{meas}},A_{\mathrm{meas}}) = 36.61\pm2.
\end{equation}
The details of the SNR-comparison measurement are displayed in table \ref{table2}. 
$S^2$ represents the power in the heterodyne frequency bin with a signal present, while $\sigma^2$ represents the variance in the same frequency bin with the same resolution bandwidth, both in units of $V^2$. 
A small increase in the amplitude between $L_{\mathrm{inj}}$ and $L_S$ on PD3 from the two laser setup measurement to the one laser setup measurement of $~3\%$ was observed, suggesting either an increase in spatial overlap between $L_S$ and the cavity eigenmode or an increase in the power of either laser. 
Since the origin of the change in amplitude is unclear, this observation is incorporated into the systematic uncertainty of the measurement. 
Other systematics derive from small changes in monitored clipping losses and power fluctuations.

A direct measurement of the cavity reflectivity was performed by comparing beatnote amplitudes between $L_{S}$ and $L_{\mathrm{refl}}$, with $L_{\mathrm{refl}}$ resonant and not resonant in the cavity on PD1. 
This gave 
$
    R_{\mathrm{cav}}(A_{\mathrm{meas}}) = \left(139\,\mu\mathrm{V}/878\,\mu\mathrm{V}\right)^2 = 0.0252
$
and is within the error of the previously calculated value in table \ref{table1}.

The directly measured $G_{\mathrm{meas}}(A_{\mathrm{meas}},A_{\mathrm{meas}})$ matches well with the calculated value $G_{\mathrm{TN}}(A_{\mathrm{meas}},A_{\mathrm{meas}})$. 
This is to be expected since the system's noise was dominated by the electronic noise of the photodetector chain during the first science campaign, as mentioned earlier 
\cite{opticsresults}. 

\begin{table}[ht]
    \centering
    \begin{tabular}{@{}lll@{}}
        \toprule
        & $S^2[\mathrm{V}^2]$ & $\sigma^2[\mathrm{V}^2]$\\
        \midrule
        1L & $2.073e-8$ & $1.422e-12$\\ 
        2L & $7.700e-7$ & $1.442e-12$\\ 
        \midrule
        2L/1L& $37.13$ & $1.014$  \\
        \midrule
        $G_{\mathrm{meas}}$ & $36.61\pm2$ &\\
        \botrule
    \end{tabular}
    \caption{\label{table2}
        Results of the individual one and two laser measurements (1L and 2L) are given in units of power spectral density. The ratios are multiplied to acquire the improvement in sensitivity of the two laser setup over the one laser setup $G$ at the total round trip attenuation $A_{\mathrm{meas}}$.
    }
\end{table}

\section{Conclusion}

In this work, we present an upgrade to the ALPS II heterodyne detection system, which utilizes a second, phase-locked, non-resonant laser incident on the experiment's regeneration cavity.
This setup allows the full power of the laser to be reflected from the cavity, thus increasing the optical power of the heterodyne local oscillator.
This upgrade replaces the previous design, in which the signal-to-noise ratio was strongly limited by the cavity in-mode reflectivity. 
Decoupling the sensitivity of the detection system from the on-resonance reflectivity of the cavity is therefore critical. This is because the excess optical losses must be reduced to a minimum to maximize the regenerated axion signal strength, but in the one laser setup doing so will also reduce the power of the laser used as the heterodyne local oscillator.

The ALPS II experiment has successfully concluded its first science campaign using this upgrade to the detection system \cite{sciresults,opticsresults}. 
Under conditions close to the science campaign, the noise model described in Section~\ref{Sec:SNR_model} predicted a factor of approximately $35 \pm4$  improvement in the SNR due to these changes in the design, which was confirmed by independent measurements described in Sections~\ref{meas} and \ref{Sec:Results}. 
Although the first science campaign was limited by stray light rather than technical noise, the upgrade was critical to reaching a sensitivity in which the stray light could be studied with far more precision than would have been possible using the previous configuration.

Further upgrades of the cavity optics and detector system electronics, in combination with the design changes presented in this work, will allow the experiment to reach even better sensitivities and perform measurements close to the quantum noise limit. 
For a given set of excess optical losses in the regeneration cavity, the highest possible sensitivity to the conversion and reconversion of axions is achieved when the cavity is impedance matched and the input transmissivity is tuned to be equal to the sum of the excess losses. 
In this case, it is only possible to operate the experiment with a second laser used in the heterodyne detection system, as the one laser system would not measure any signal at all due to the on-resonance reflectivity of the cavity being zero.

\backmatter

\bmhead{Abbreviations}
ALPS II: Any light particle search II; RC: regeneration cavity; HOM: Higher-order mode; SNR: Signal-to-noise ratio; PDH: Pound-Drever-Hall; PLL: Phase-locked loop


\bmhead{Acknowledgments}
The Authors would like to thank the members of the ALPS collaboration for their support, and especially Daniel C. Brotherton, for his help automating the system.

\bmhead{Author's contributions}
HF and AS conceived the study and developed the methodology. HF performed the investigation and analysis, curated and interpreted the data, prepared the figures, and wrote the original draft of the manuscript. AS, AL, and GM supervised the work, provided guidance on the study design and interpretation of the results, and reviewed the manuscript. All authors read and approved the final manuscript.

\bmhead{Funding}
We acknowledge the support of the National Science Foundation (Grant 
No.\,1802006), of the Heising-Simons Foundation (Grant No.\,2015-154 and
2020-1841), of the Deutsche Forschungsgemeinschaft through project grant
WI 1643/2-1 and EXC 2121 ``Quantum Universe'' – 390833306, of the
Science and Technology Facilities Council (UK), grants ST/T006331/1 and
ST/Y004515/1, as well as support by the German Volks\-wagen Stiftung and
the European Research Council (ERC) under the European Union’s Horizon
2020 research and innovation program Grant agreement No.\,948689.

\bmhead{Availability of data and materials}
The datasets used and/or analyzed during the current study are available upon reasonable request.

\bmhead{Competing interests}
The authors declare that they have no competing interests.

\bibliography{sn-bibliography}

\end{document}